# Line, LINER, linest - from micro-AGN to ultra-luminous LINERs. One and the same?

**Hartmut Winkler**

Department of Physics, University of Johannesburg, PO Box 524, 2006 Auckland Park, Johannesburg, South Africa

E-mail: hwinkler@uj.ac.za

**Abstract.** This paper compares the optical spectra of a wide range of galaxies categorised as members of the Low Ionisation Nuclear Emission Region (LINER) class of active galactic nuclei (AGN). LINERs are defined by emission spectra with relatively faint high ionisation lines (compared to other AGN classes). The gas emission luminosity ranges from the weak flux emanating from some nearby galactic nuclei all the way to extremely luminous radio galaxies, where the line emission can completely dominate the host galaxy starlight component. In this study I analyse the Sloane Digital Sky Survey optical spectra of 15 LINERS identified in the course of the preparation of the new edition of the ZORROASTER AGN catalogue, spanning the largest possible luminosity range. I compare relative emission line strengths, focusing on uncommonly analysed ratios such as those involving [N I], line widths, profiles and even the spectral features of the host galaxy stellar continuum. The study identifies possible luminosity-dependent trends in the spectral properties of the studied objects. Possible reasons are presented to rationalise these trends, and the paper concludes with a discussion regarding the uniformity of the LINER class.

## 1. Introduction

The second version of the ZORROASTER[1] online catalogue of optical spectra of active galactic nuclei (AGN), to be released in October 2013, will now include numerous narrow-line objects. In particular, the update contains a significant number of entries meeting the general criteria of a class of AGN referred to as LINER (Low Ionisation Nuclear Emission-line Region).

The LINER class was originally suggested by Heckman [1], and its main characteristics are the comparatively strong spectral lines corresponding to low ionisation features such as [O I] and [N I]. Conversely, the high excitation [O III] lines are much weaker than found in most other types of AGN (e.g. Seyferts and starburst galaxies).

It is not clear that LINERs constitute a homogeneous class of objects. Various mechanisms have been suggested to explain the general features of the optical spectra, including i) photoionisation by an accretion disk surrounding a black hole (as in Seyfert galaxies), ii) photoionisation by stars (as in starburst galaxies) and iii) shock heated gas (as in some radio galaxies) [2,3,4]. None of these models have been completely discounted for all objects associated with this classification. It should hence be considered that several classes of AGN share similar spectral features, but are in fact quite disparate types of objects with very different physical characteristics.

---

[1] http://www.uj.ac.za/EN/Faculties/science/departments/physics/

It has become common to use line ratios as the defining parameter to distinguish LINERs from alternative AGN classes. In particular, two-dimensional logarithmic plots of one line ratio versus another offer a convenient systematic tool to automatically assign activity classes to large samples of galaxies [5,6]. However, by failing to carefully consider the peculiarities and full details of individual objects, this procedure is prone to misclassifications.

The line ratio based classification scheme also does not consider the wide range observed in LINER luminosities. It is well established that a lot of the otherwise unremarkable nuclei of nearby galaxies exhibit faint LINER emission spectra [7]. But highly luminous AGN like Arp 102B are also viewed as part of this class [8]. The luminosity dependence of LINER spectra is not properly established.

This paper investigates the spectra of 15 objects that satisfy the general LINER classification requirements according to the original definition [1], and explores possible line ratio vs. luminosity correlations. Their coordinates, redshifts and common names (where applicable) are listed in table 1.

**Table 1.** LINERs studied in this paper.

| RA(2000) | Dec(2000) | $z$ | common name |
|----------|-----------|-----|-------------|
| 02h45m27.5s | +00°54′52″ | 0.0244 | Mkn 1052 |
| 03h16m54.9s | −00°02′31″ | 0.0232 | Mkn 1074 |
| 03h19m48.1s | +41°30′42″ | 0.0176 | NGC 1275 |
| 09h22m05.2s | +00°09′03″ | 0.0694 | |
| 09h33m46.1s | +10°09′09″ | 0.0108 | NGC 2911 |
| 09h39m17.2s | +36°33′44″ | 0.0197 | |
| 09h43m19.2s | +36°14′52″ | 0.0221 | NGC 2965 |
| 09h50m50.2s | +09°52′37″ | 0.0512 | |
| 10h01m31.2s | +46°59′46″ | 0.0860 | |
| 12h15m00.8s | +05°00′51″ | 0.0782 | |
| 15h08m42.8s | +03°33′10″ | 0.0549 | |
| 15h24m12.6s | +08°32′40″ | 0.0371 | |
| 15h26m06.2s | +41°40′14″ | 0.0083 | NGC 5929 |
| 16h52m58.9s | +02°24′03″ | 0.0245 | NGC 6240 |
| 21h25m12.5s | −07°13′30″ | 0.0639 | |

## 2. Spectral data analysis

### 2.1. Archival spectra

For the majority of objects analysed here, electronic spectra are available in the Sloan Digital Sky Survey (SDSS) database [9]. These cover the entire optical range, and successive data point in the spectrum corresponds to adjacent channels of wavelength intervals of $\Delta(\log \lambda) = 0.0001$. In addition to the SDSS objects, two well-studied AGN with LINER features (NGC 1275 and NGC 6240) were included in the sample as well. Archival spectra of these were also available in electronic form [10,11], and were converted into the SDSS format. All spectra are displayed in figure 1.

### 2.2. Bands

The 2013 version of the ZORROASTER database is set to include total fluxes for selected bands consisting of 11 adjacent channels. These are strategically chosen to include all the major emission features in AGN, including the following: [O II] 3727 Å, [O III] 5007 Å, [N II] 5198 Å, [O I] 6300 Å, [N II] 6584 Å and [S II] 6728 Å. Further bands are defined on either side of these lines, allowing continuum subtraction. In that way the line flux can be estimated in a crude but robust manner that makes no specific assumptions regarding line widths and profiles.

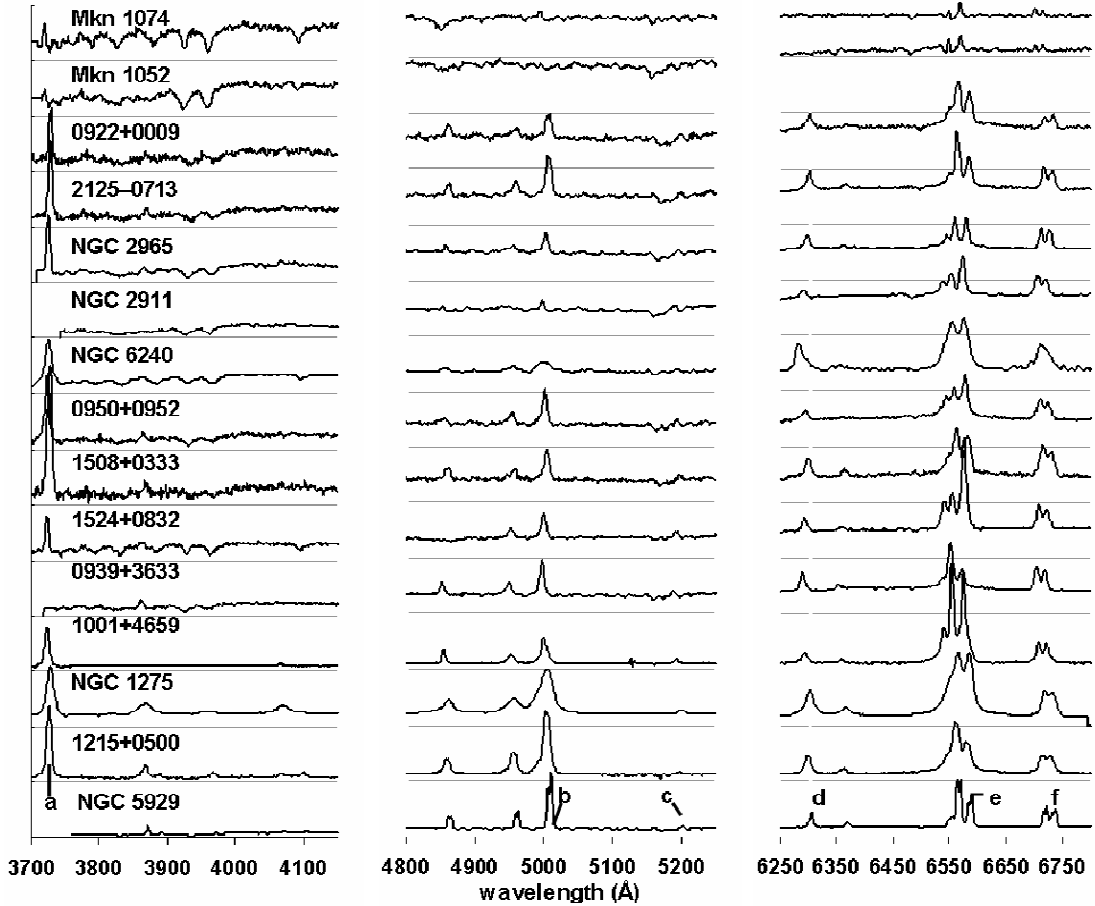

**Figure 1.** Sections of the redshift corrected spectra of the LINERs investigated in this study including the emission lines utilised in this study, whose positions are marked as follows: a - [O II] 3727 Å, b - [O III] 5007 Å, c - [N I] 5198 Å, d - [O I] 6300 Å, e - [N II] 6584 Å, f - [S II] 6725 Å.

## 3. Results

The comparative emission line region luminosity $L$ was parameterised through multiplication of the continuum-subtracted band fluxes by $z^2$. The second column of table 2 gives the values of $\log L$ (plus an arbitrary constant) for the band flux corresponding to [N II] 6584 Å. The remaining columns of table 2 quantify the ratio of the continuum-corrected band fluxes of the earlier mentioned lines representative of [N I], [N II], [O I], [O II] and [S II] relative to [O III] 5007 Å. These are presented in the form

$$R([x]/[y]) = \log\big(f([x])/f([y])\big) .$$

A variety of plots were constructed to test the interdependence between line ratios and the emission line region luminosity for the different elements and states of ionisation. In each case $R([x]/[y])$ was plotted against $\log L$ for the combinations listed in table 3. The respective slopes and degrees of correlation (in terms of the standard correlation coefficient squared) obtained are given in that table.

**Table 2.** Relative distance corrected luminosity of the emission line region (determined from [N II] 6583 Å) and line ratios (given as $R([x]/[y])$). Due to the weakness and resultant unreliability of its measured line strengths, the results for Mkn 1052 are not included here.

| Name | log $L$([N II]) + C | $R$([N I]/ [O III]) | $R$([N II]/ [O III]) | $R$([O I]/ [O III]) | $R$([O II]/ [O III]) | $R$([S II]/ [O III]) |
|---|---|---|---|---|---|---|
| Mkn 1074 | 0.38 | | 0.33 | −0.16 | 0.43 | 0.38 |
| NGC 1275 | 2.08 | −1.08 | 0.09 | −0.22 | −0.10 | 0.12 |
| SDSS 0922+0009 | 1.73 | | 0.30 | −0.09 | 0.23 | 0.08 |
| NGC 2911 | 1.31 | −0.56 | 1.07 | 0.08 | | 0.83 |
| SDSS 0939+3633 | 1.27 | −1.47 | −0.07 | −0.22 | | 0.20 |
| NGC 2965 | 1.26 | | 0.27 | −0.05 | 0.43 | 0.40 |
| SDSS 0950+0952 | 1.83 | −1.07 | 0.31 | −0.41 | 0.23 | 0.22 |
| SDSS 1001+4659 | 2.78 | −0.86 | 0.58 | −0.35 | 0.08 | 0.22 |
| SDSS 1215+0500 | 2.43 | −1.41 | −0.30 | −0.48 | −0.10 | −0.18 |
| SDSS 1508+0333 | 1.60 | | 0.32 | −0.04 | 0.62 | 0.47 |
| SDSS 1524+0832 | 1.76 | −0.81 | 0.46 | −0.23 | 0.08 | 0.37 |
| NGC 5929 | 1.27 | −1.40 | −0.09 | −0.52 | | −0.02 |
| NGC 6240 | 1.78 | −0.73 | 0.84 | 0.20 | 0.51 | 0.70 |
| SDSS 2125−0713 | 1.69 | −1.25 | 0.01 | −0.26 | 0.16 | 0.11 |

**Table 3.** Correlation tests for the line ratio vs. line luminosity graphs. The entries correspond to the slope and (in brackets) $R^2$ correlation parameter. Bold entries correspond to the strongest correlations, significant at the 95% level.

| | $R$([N II]/ [N I]) | $R$([O II]/ [O I]) | $R$([N I]/ [O III]) | $R$([N II]/ [O III]) | $R$([O I]/ [O III]) | $R$([O II]/ [O III]) | $R$([S II]/ [O III]) |
|---|---|---|---|---|---|---|---|
| log $L$([N I]) | −0.162 (0.326) | −0.033 (0.009) | −0.063 (0.013) | −0.226 (0.086) | −0.176 (0.193) | **−0.339** **(0.539)** | −0.285 (0.283) |
| log $L$([N II]) | −0.108 (0.109) | −0.105 (0.157) | +0.055 (0.008) | −0.039 (0.004) | −0.107 (0.088) | **−0.239** **(0.368)** | −0.162 (0.117) |
| log $L$([O I]) | −0.202 **(0.465)** | −0.122 (0.205) | −0.226 (0.157) | −0.265 (0.184) | −0.132 (0.147) | −0.256 **(0.413)** | −0.268 **(0.345)** |
| log $L$([O II]) | −0.168 (0.222) | −0.060 (0.041) | −0.370 **(0.448)** | −0.226 (0.176) | −0.171 (0.239) | −0.230 (0.278) | −0.253 (0.349) |
| log $L$([S II]) | −0.210 (0.349) | −0.106 (0.133) | −0.174 (0.064) | −0.212 (0.093) | −0.150 (0.150) | −0.265 **(0.378)** | −0.244 (0.226) |

Examples of some of these plots are shown in figures 2 and 3.

## 4. Discussion

Even a cursory visual inspection of the spectra displayed in figure 1 reveals peculiar characteristics for practically every one of the objects analysed, even though they are ostensibly all of the same AGN class. Mkn 1074 has a continuum with comparatively deep hydrogen absorption features, indicating a relatively high proportion of early type stars. Both Markarian objects only have very weak emission features. SDSS 1001+4659 appears to have narrow starburst-like peaks superimposed on some emission lines, while several other objects (e.g. the well-known NGC 1275) display unusually wide forbidden lines. The [O I] line strengths in some objects (e.g. NGC 6240) are far greater than elsewhere (e.g. NGC 2911).

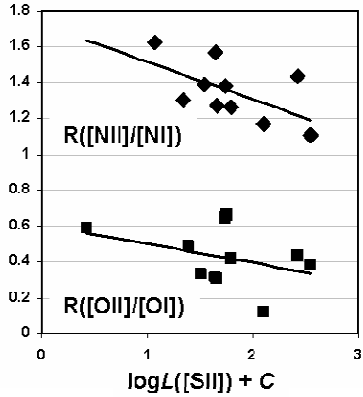

**Figure 2.** Plot of some line ratios vs. the [S II] emission line region luminosities

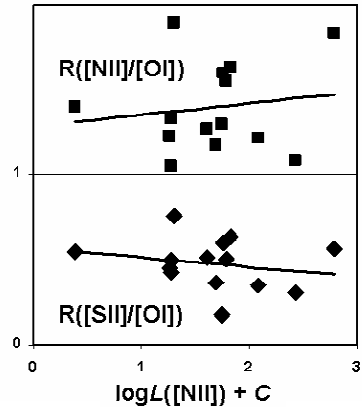

**Figure 3.** Plot of some line ratios vs. the [N II] emission line region luminosities.

The fact that there seems to be a dependence of the line ratios on luminosity is not entirely surprising, as both the plotted parameters depend on the strength and spectral distribution of the excitational radiation and element abundances. It is however notable that some ratios, for example $R$([O II]/[O III]) and $R$([S II]/[O III]), correlate far better with luminosities than, say, R([O II]/[O I]). Also, the [O I] region luminosity is usually much better correlated with the line ratios than the [N II] region luminosity.

The complete lack of correlation in particular cases such as those illustrated in figure 3 suggests that these may be indicative of intrinsically different line forming mechanisms in some of the objects plotted here. These may paradoxically be optimal for differentiating between different sub-classes of LINERs undergoing distinct physical processes.

## 5. Conclusion

The study confirms that what we collectively refer to as LINERs are likely to be an inhomogeneous group of AGN, generated by a variety of very disparate processes. Specific plots of line ratios versus emission line region luminosities show promise of differentiating between the various types of objects. The achievement of this will however require the analysis of a vastly larger sample of LINERs, something that will soon be possible due to the imminent expansion of the ZORROASTER database.

## Acknowledgments

This paper utilized data from the Sloan Digital Sky Survey (SDSS). Funding for the SDSS and SDSS-II has been provided by the Alfred P. Sloan Foundation, the Participating Institutions, the National Science Foundation, the U.S. Department of Energy, the National Aeronautics and Space Administration, the Japanese Monbukagakusho, the Max Planck Society, and the Higher Education Funding Council for England. The SDSS Web Site is http://www.sdss.org/.

The SDSS is managed by the Astrophysical Research Consortium for the Participating Institutions. The Participating Institutions are the American Museum of Natural History, Astrophysical Institute Potsdam, University of Basel, University of Cambridge, Case Western Reserve University, University of Chicago, Drexel University, Fermilab, the Institute for Advanced Study, the Japan Participation Group, Johns Hopkins University, the Joint Institute for Nuclear Astrophysics, the Kavli Institute for Particle Astrophysics and Cosmology, the Korean Scientist Group, the Chinese Academy of Sciences

(LAMOST), Los Alamos National Laboratory, the Max-Planck-Institute for Astronomy (MPIA), the Max-Planck-Institute for Astrophysics (MPA), New Mexico State University, Ohio State University, University of Pittsburgh, University of Portsmouth, Princeton University, the United States Naval Observatory, and the University of Washington.

## References

[1]   Heckman T M 1980 *Astron. & Astrophys.* **87** 152
[2]   Ho L C, Filippenko A V and Sargent W L W 1993 *Astrophys. J.* **417** 63
[3]   Filippenko A V 2003 LINERs and their physical mechanisms *ASP Conf. Series* **290** 369
[4]   Dopita M A and Sutherland R S 1995 *Astrophys. J.* **455** 468
[5]   Veilleux S and Osterbrock D E 1987 *Astrophys. J. Suppl.* **63** 295
[6]   Kewley L J, Groves B, Kauffmann G and Heckman T 2006 *Monthly Notices Roy. Astron. Soc.* **372** 961
[7]   Ho L C, Filippenko A V and Sargent W L W 1997 *Astrophys. J. Suppl.* **112** 312
[8]   Halpern J P, Eracleous M, Filippenko A V and Chen K 1996 *Astrophys. J.* **464** 704
[9]   Abazajian K, Adelman J, Agueros M, et al. 2003 *Astron. J.* **126** 2081
[10]  Kennicutt R C 1992 *Astrophys. J. Suppl.* **79** 255
[11]  Moustakas J and Kennicutt R C 2006 *Astrophys. J. Suppl.* **164** 81